\documentclass[10pt]{article}

\begin{document}

\baselineskip=4.4mm

\makeatletter

\newcommand{\E}{\mathrm{e}\kern0.2pt}
\newcommand{\D}{\mathrm{d}\kern0.2pt}
\newcommand{\RR}{\mathrm{I\kern-0.20emR}}

\def\bottomfraction{0.9}

\title{\bf On bounds and non-existence in the problem of steady waves with
vorticity}

\author{Vladimir Kozlov$^1$, Nikolay Kuznetsov$^2$ and Evgeniy Lokharu$^1$}

\date{}

\maketitle

\vspace{-8mm}

\begin{center}
$^1$Department of Mathematics, Link\"oping University, S--581 83 Link\"oping, Sweden
\\ $^2$ Laboratory for Mathematical Modelling of Wave Phenomena, \\ Institute for
Problems in Mechanical Engineering, Russian Academy of Sciences, \\ V.O., Bol'shoy
pr. 61, St. Petersburg 199178, Russian Federation \\ E-mail: vladimir.kozlov@liu.se;
nikolay.g.kuznetsov@gmail.com; evgeniy.lokharu@liu.se
\end{center}

\begin{abstract}
For the problem describing steady, gravity waves with vorticity on a
two-dimensional, unidirectional flow of finite depth the following results are
obtained. (i)~Bounds for the free-surface profile and for Bernoulli's constant.
(ii)~If only one parallel shear flow exists for a given value of Bernoulli's
constant, then there are no wave solutions provided the vorticity distribution is
subject to a certain condition.

\vspace{1mm}


\end{abstract}

\setcounter{equation}{0}

\section{Introduction}

We consider the two-dimensional nonlinear problem describing steady waves in a
horizontal open channel of uniform rectangular cross-section. The motion of an
inviscid, incompressible, heavy fluid, say, water occupying the channel is supposed
to be rotational with a prescribed vorticity distribution. This type of motion
commonly occurs in nature as is indicated by observations (see, for example,
\cite{SCJ,Th} and references cited therein). The plethora of results obtained for
various models describing waves with vorticity is briefly characterised by
\cite{KK2}. In that paper, flows with counter-currents as well as unidirectional
ones were studied. Further details about the latter flows concerning, in particular,
global branches of solutions can be found in \S~3 of the survey article by
\cite{WS}.

Here, our aim is to investigate some new properties of the problem about waves on
unidirectional flows. In particular, we generalise fundamental bounds for steady
irrotational waves found by \cite{KK1} (see also \cite{KN1} and \cite{AT} for the
cases of Stokes and solitary waves, respectively), thus extending results obtained
in \cite{KK0}.

\subsection{Statement of the problem}

Let an open channel of uniform rectangular cross-section be bounded below by a
horizontal rigid bottom and let water occupying the channel be bounded above by a
free surface not touching the bottom. The water motion is supposed to be
two-dimensional and rotational; the surface tension is neglected on the free surface
of water, where the pressure is constant. These assumptions and the fact that water
is incompressible allow us to seek the velocity field in the form $(\psi_y,
-\psi_x)$, where $\psi (x,y)$ is referred to as the {\it stream function}. The
vorticity distribution $\omega$ is supposed to be a prescribed continuous function
depending on $\psi$.

The non-dimensional variables proposed by \cite{KN} (see Appendix~A in \cite{KK2}
for details of scaling) are used here. In particular, lengths and velocities are
scaled to $(Q^2/g)^{1/3}$ (the depth of the critical uniform stream in the
irrotational case) and $(Qg)^{1/3}$, respectively; $Q$ and $g$ are the dimensional
quantities for the rate of flow and the gravity acceleration, respectively.

In appropriate Cartesian coordinates $(x,y)$, the bottom coincides with the $x$-axis
and gravity acts in the negative $y$-direction. We choose the frame of reference so
that the velocity field is time-independent as well as the unknown free-surface
profile. The latter is assumed to be the graph of $y = \eta (x)$, $x\in \RR$, where
$\eta$ is a positive continuous function, and so the longitudinal section of the
water domain is $D = \{ x \in \RR,\ 0 < y < \eta (x) \}$.

The following free-boundary problem for $\psi$ and $\eta$ that describes all kinds
of waves has long been known (cf. \cite{KN}):
\begin{eqnarray}
&& \psi_{xx} + \psi_{yy} + \omega (\psi) = 0, \quad (x,y)\in D; \label{eq:lapp} \\ &&
\psi (x,0) = 0, \quad x \in \RR; \label{eq:bcp} \\ && \psi (x,\eta (x)) = 1, \quad x
\in \RR; \label{eq:kcp} \\ && |\nabla \psi (x,\eta (x))|^2 + 2 \eta (x) = 3 r, \quad x
\in \RR . \label{eq:bep}
\end{eqnarray}
In condition (\ref{eq:bep}) (Bernoulli's equation), $r$ is a constant considered as
the problem's parameter and referred to as Bernoulli's constant/the total head. In
what follows, we suppose that $\psi$ is a monotonic function of $y$ for all $x \in
\RR$ because we are going to study unidirectional flows.

\subsection{Main results}

Prior to formulating our results we list some auxiliary facts required in what
follows.

By a {\it stream} (shear-flow) solution of problem (\ref{eq:lapp})--(\ref{eq:bep})
we mean a pair $(u (y), \, d)$ (the constant depth of flow is given by $d$) such
that the following relations hold:
\begin{equation} 
u'' + \omega (u) = 0 \ \ \mbox{on} \ (0, d) , \ \ \ u (0) = 0, \ \ \ u (d) = 1 , \
\ \ |u' (d)|^2 + 2 \, d = 3 \, r , \label{eq:ss1}
\end{equation}
here the prime denotes the differentiation with respect to $y$. A detailed study of
these solutions, in particular, those describing flows with counter-currents is
given in \cite{KK3}. The set of unidirectional solutions of the first three
relations (\ref{eq:ss1}) is parameterised by $s = u' (0)$; it is greater than or
equal to $s_0 = \sqrt{2 \max_{0 \leq \tau \leq 1} \Omega (\tau)}$ as follows from
the following expressions for $u$ and $d$ (implicit and explicit, respectively):
\begin{equation}
y = \int_0^u \frac{\D \tau}{\sqrt{s^2 - 2 \, \Omega (\tau)}}  \ \ \mbox{and} \ \ d =
\int_0^1 \frac{\D \tau}{\sqrt{s^2 - 2 \, \Omega (\tau)}} , \ \ \mbox{where} \ \Omega
(\tau) = \int_0^\tau \omega (t) \, \D t \, . \label{eq:d}
\end{equation}
(Note that $\Omega$ belongs to $C^1 ([0, 1])$.) It is clear that $d \, [= d (s)]$
decreases strictly monotonically and tends to zero as $s \to +\infty$, whereas $d_0
= \lim_{s \to s_0 + 0} d (s)$ can be finite or infinite depending on the behaviour
of $\Omega$ on $[0, 1]$ (see below).

The first formula (\ref{eq:d}) and last relation (\ref{eq:ss1}) imply the equation
\begin{equation}
r = {\cal R} (s) , \quad \mbox{where} \ {\cal R} (s) = [ s^2 - 2 \, \Omega (1) + 2
\, d (s) ] / 3 . \label{eq:calR}
\end{equation}
It is clear that this function has only one minimum, say, $r_c > 0$ attained at some
$s_c > s_0$. If $d_0 = +\infty$ and $r > r_c$, then (\ref{eq:calR}) has two
solutions $s_+$ and $s_-$ such that $s_0 < s_+ < s_c < s_-$. Substituting $s_+$ and
$s_-$ into (\ref{eq:d}), one obtains the stream solutions $(u_+, d_+)$ and $(u_-,
d_-)$, respectively. The shear flows described by these solutions are analogous to
the uniform sub- and supercritical flows existing in the irrotational case. If $d_0
< +\infty$, then both $s_+$ and $s_-$, and consequently, the corresponding stream
solutions exist only for $r \in (r_c, r_0)$, where $r_0 = {\cal R} (s_0)$. If $r >
r_0$, then only $s_-$ exists and defines $(u_-, d_-)$.

In order to describe how $d_0$ and the corresponding stream solution depend on the
vorticity distribution \cite{KK3} considered the following three options:

\vspace{1mm}

\noindent \ \ (i) $\max_{0 \leq \tau \leq 1} \Omega (\tau)$ is attained either at an
inner point of $(0, 1)$ or at one (or both) of the end-points. In the latter case,
either $\omega (1) = 0$ when $\Omega (1) > \Omega (\tau)$ for $\tau \in (0, 1)$ or
$\omega (0) = 0$ when $\Omega (0) > \Omega (\tau)$ for $\tau \in (0, 1)$ (or both of
these conditions hold simultaneously).

\vspace{1mm}

\noindent \ (ii) $\Omega (0) > \Omega (\tau)$ for $\tau \in (0, 1]$ and $\omega (0)
< 0$.

\vspace{1mm}

\noindent (iii) $\Omega (\tau) < \Omega (1)$ for $\tau \in (0, 1)$ and $\omega (1) >
0$. Moreover, if $\Omega (1) = 0$, then $\omega (0) < 0$ and $\omega (1) > 0$ must
hold simultaneously.

\vspace{1mm}

\noindent Conditions (i)--(iii) define three disjoint sets of vorticity
distributions whose union is the whole set of distributions continuous on $[0,1]$.
It occurs that conditions (i) imply that $d_0 = +\infty$, whereas conditions (ii)
and (iii) yield that $d_0$ is finite. The last two conditions have the following
consequences for solutions of problem (\ref{eq:ss1}): $u' (d_0) = 0$ under
conditions (iii), whereas (ii) implies that $u' (0) = 0$ and $u' (d_0) \neq 0$.
 
In order to formulate our result about fundamental bounds for $\hat \eta = \sup_{x
\in \RR} \eta (x)$ and $\check \eta = \inf_{x \in \RR} \eta (x)$ we have to impose
some restrictions on $\psi$ and $\eta$.

\vspace{1mm}

\noindent {\sc Conditions (A).} The function $\psi$ belongs to $C^{1, \alpha} (\bar
D)$; that is, $|\psi|$, $|\psi_x|$ and $|\psi_y|$ are bounded on $\bar D$, whereas
the derivatives satisfy the H\"older condition there. Moreover, $\psi_y (x, y) \geq
\delta$ on $\bar D$ for some $\delta > 0$. The function $\eta$ belongs to $C^{0,
\alpha}_{loc} (\RR)$.

\vspace{1mm}

\noindent {\sc Theorem 1.} {\it Let for some $r > 0$ problem
$(\ref{eq:lapp})$--$(\ref{eq:bep})$ have a non-stream solution $(\psi, \eta)$ satisfying
conditions {\rm (A)}. Then the following two assertions hold.

$(1)$ $r > r_c$ and $\eta (x) > d_-$ for all $x \in \RR$.

$(2)$ If $r$ belongs to $(r_c , r_0)$, then the inequalities $\hat \eta \geq d_+ >
\check \eta$ are true. Moreover, the left inequality is also strict provided $\hat
\eta$ is attained at some point.

Furthermore, if $\omega$ satisfies conditions {\rm (iii)}, then every solution of
problem $(\ref{eq:lapp})$--$(\ref{eq:bep})$ satisfying conditions {\rm (A)} is a stream
solutions provided $r \geq r_0$.}

\vspace{1mm}

We expect that the left inequality in assertion (2) is always strict which is the
case for irrotational waves; see \cite{KK1}, where assertions similar to (1) and (2)
were obtained. However, there is no analogue of the last assertion for irrotational
waves. On the other hand, it extends the result of \cite{KK4} about the absence of
small-amplitude non-stream solutions (not necessarily unidirectional) when $r =
r_0$, $s = s_0 > 0$ and a certain restriction is imposed on vorticity. Besides,
Theorem~1 does not cover the case when $r > r_0$ and the vorticity distribution
satisfies conditions (ii) and we discuss it in \S~2.3.

\subsection{The partial hodograph transform}

Since we consider unidirectional flows, it is convenient to reformulate problem
(\ref{eq:lapp})--(\ref{eq:bep}) using the partial hodograph transform; that is,
mapping the unknown domain $D$ onto the strip $S = \RR \times (0, 1)$ (cf. \cite{GW}
and \cite{CS}):
\[ \bar D \ni (x,y) \mapsto (q,p) \in \bar S , \ \ \mbox{where} \ q = x \ \mbox{and}
\ p = \psi (x,y) .
\]
These variables are treated as independent, whereas $y = h (q,p)$ is the new unknown
function such that $h_p (q,p) \geq \delta' > 0$ on $\bar S$ which follows from
conditions (A). A straightforward calculation shows that problem
(\ref{eq:lapp})--(\ref{eq:bep}) takes the following form:
\begin{equation}
{\cal D} h = 0 , \ (q,p) \in S ; \ \ h (q,0) = 0, \ q \in \RR ; \ \
(1+h_q^2) h_p^{-2} + 2 h = 3 r, \ p = 1 , \ q \in \RR . \label{eq:9}
\end{equation}
Here ${\cal D} h$ stands for
\[ \bigg[ \frac{h_q}{h_p} \bigg]_q - \bigg[ \frac{1+h_q^2}{2 h_p^2} +
\Omega (p) \bigg]_p .
\]
On the other hand, a solution of this problem allows us to recover $\eta$ using the
equality
\begin{equation}
\eta (x) = h(x, 1) , \quad x \in \RR . \label{eq:reco}
\end{equation}
Therefore, we write $\hat \eta$ and $\check \eta$ for $\sup_{q \in \RR} h (q, 1)$
and $\inf_{q \in \RR} h (q, 1)$, respectively, in what follows. Note that $h (q, 1)
> 0$ because $h(q, 0) = 0$ and $h_p$ is positive in $S$. In order to recover $\psi$
one has to solve the Dirichlet problem (\ref{eq:lapp})--(\ref{eq:kcp}) in the domain
whose upper boundary is defined by (\ref{eq:reco}).

\section{Proof of Theorem 1} 

To prove Theorem 1 we reformulate it in terms of problem (\ref{eq:9}).

\vspace{1mm}

\noindent {\sc Theorem $1'$.} {\it Let for some $r > 0$ problem $(\ref{eq:9})$ have a
non-stream solution $h \in C^{1, \alpha} (\bar S)$ such that $h_p \geq \delta'$ on
$\bar S$ for some $\delta' > 0$. Then the following two assertions hold.

$(1)$ $r > r_c$ and $h (q, 1) > d_-$ for all $q \in \RR$.

$(2)$ If $r$ belongs to $(r_c , r_0)$, then the inequalities $\hat \eta \geq d_+ >
\check \eta$ are true.

Moreover, if $\omega$ satisfies conditions {\rm (iii)} and $r \geq r_0$, then every
solution of problem $(\ref{eq:9})$ belonging to $C^{1, \alpha} (\bar S)$ is a stream
solution.}

\subsection{Auxiliary propositions}

The proof of Theorem $1'$ is based on two auxiliary propositions. The first of these
is maximum principle for an elliptic operator that arises when considering ${\cal D}
\xi - {\cal D} \zeta$, where $\xi$ and $\zeta$ are two different solutions of the
first equation (\ref{eq:9}). In \cite{CS} (see pp.~155, 156), it is shown that $u =
\xi - \zeta$ satisfies the equation
\begin{equation}
{\cal L} u = \big[ a^{(qq)} \, u_q + a^{(qp)} \, u_p \big]_q + \big[ a^{(pp)} \, u_p
+ a^{(pq)} \, u_q \big]_p = 0 \quad \mbox{in} \ S , \label{eq:L}
\end{equation}
which is uniformly elliptic provided $\xi$ and $\zeta$ have bounded gradients and
$\xi_p$ and $\zeta_p$ are greater than some positive constant.

\vspace{1mm}

\noindent {\sc  Proposition 1.} {\it Let $u \in W^{1,2}_{loc} (\bar S)$ be bounded
in $\bar S$. If ${\cal L} u = 0$ on $S$, where ${\cal L}$ has the form $(\ref{eq:L})$
and its coefficients are bounded and such that ${\cal L}$ is uniformly elliptic on
$\bar S$ with the ellipticity constant $C_{\cal L} > 0$, then the following
equalities hold:}
\begin{equation}
\sup_{S} u (q,p) = \sup_{\RR} \{ u (q, 0) , u (q, 1) \} \ \ and \ \ \inf_{S} u (q,p)
= \inf_{\RR} \{ u (q, 0) , u (q, 1) \} .
\label{eq:pr_1}
\end{equation}

{\it Proof.} It is clear that the second equality follows from the first one. Note
that the assumptions imposed on $u$ imply its H\"older continuity (see, for example,
\S~8.9 in \cite{GT}). Moreover, the H\"older norm of $u$ over $[t, t+1] \times [0,
1]$ is bounded by a constant depending on $C_{\cal L}$, the H\"older exponent and
various bounds (for the coefficients of ${\cal L}$ and for $u$ itself), but
independent of $t$.

In order to prove the first equality (\ref{eq:pr_1}) we assume the contrary; that
is, $\hat u = \sup_{S} u$ is strictly greater than $\sup_{q \in \RR} \{ u (q, 0) , u
(q, 1) \}$. Then there exist positive numbers $\epsilon$ and $\ell$ (the latter
depends on $\epsilon$, generally speaking) such that
\begin{equation}
\hat u - u (q,p) \geq \epsilon \quad \mbox{on} \quad S_\ell = \{ (q,p) \in S : q \in
\RR, \ p \in (0, \ell) \cup (1 - \ell, 1) \} .
\label{eq:ell}
\end{equation}
It also exists a sequence $\{ (q_k,p_k) \}_{k=1}^\infty$ such that $\ell < p_k < 1 -
\ell$ and $u (q_k,p_k) \to \hat u$.

Let $B_\rho (q,p)$ denote the open circle of radius $\rho$ centred at $(q,p)$. Since
$\hat u - u \geq 0$ satisfies ${\cal L} (\hat u - u) = 0$ in $S$, Harnack's
inequality (see Corollary~8.21 in \cite{GT}) is applicable in every $B_{\rho
(\ell)} (q_k,1/2)$, where $\rho (\ell) = (1-\ell)/2$, and so $(q_k,p_k) \in B_{\rho
(\ell)} (q_k,1/2)$. Therefore, we have
\[ \sup_{B_{\rho (\ell)} (q_k,1/2)} (\hat u - u) \leq C 
\inf_{B_{\rho (\ell)} (q_k,1/2)} (\hat u - u) \leq C \big[ \hat u - u (q_k,p_k)
\big] \to 0 \ \ \mbox{as} \ k \to \infty
\]
for some $C > 0$ that depends on bounds for the coefficients of ${\cal L}$, $C_{\cal
L}$ and $\ell$. Hence the supremum on the left is arbitrarily small provided $k$ is
sufficiently large, but this is incompatible with (\ref{eq:ell}), because $S_\ell$
overlaps with $B_{\rho (\ell)} (q_k,1/2)$ for all $k=1,\dots,\infty$. The obtained
contradiction proves the proposition.

It is straightforward to verify that for any $s > s_0$ such that ${\cal R} (s) = r$
the corresponding stream solution of problem (\ref{eq:9}), say, $H (p; s)$ has the
following form:
\begin{equation}
H(p; s) = \int_{0}^{p} \frac {\D \tau}{\sqrt{s^2 - 2 \Omega (\tau)}} \, .
\label{eq:10}
\end{equation}
If $s = s_0$ and either of conditions (ii), (iii) is fulfilled, then this formula
gives a continuous function whose derivative is infinite at one of the end-points of
$[0, 1]$.

The next proposition provides an ersatz of Hopf's lemma for an elliptic equation of
the form (\ref{eq:L}) in $\bar S$.

\vspace{1mm}

\noindent {\sc  Proposition 2.} {\it Let ${\cal L}$ be an elliptic operator of the
form $(\ref{eq:L})$. If $u \in C^{1, \alpha} (\bar S)$ satisfies the equation ${\cal
L} u = 0$ in $S$ and $\sup_S u = \sup_{q \in \RR} u(q,1)$, then for any sequence $\{
q_k \}_{k=1}^\infty$ such that
\begin{equation}
u (q_k, 1) \to \sup_{q \in \RR} u (q, 1) \quad as \ k \to \infty \label{eq:11}
\end{equation}
the following relations hold:
\begin{equation}
u_q (q_k, 1) \to 0 \ \ as \ k \to \infty \quad and \quad {\rm lim\,sup}_{k \to
\infty} \ u_p (q_k, 1) \geq 0 . \label{eq:12}
\end{equation}
The similar assertion is true with supremum changed to infimum in the assumption,
whereas {\rm lim\,sup} is replaced by {\rm lim\,inf} and the inequality sign is
opposite in the second relation $(\ref{eq:12})$.

If supremum is attained at some point $(q_*, 1)$, then $u_p (q_*, 1) > 0;$ if
infimum is attained, the inequality sign is opposite.}

\vspace{1mm}

{\it Proof.} The last assertion for supremum is a consequence of Proposition~1 and
Hopf's lemma, whereas for infimum it follows by changing $u$ to $-u$. Therefore, we
assume that $\hat u = \sup_S u = \sup_{q \in \RR} u(q,1)$ and take any sequence $\{
q_k \}_{k=1}^\infty$ satisfying (\ref{eq:11}) (without loss of generality, we take
it tending to $+\infty$).

In order to prove relations (\ref{eq:12}) we assume the contrary, that is, there
exist $\epsilon > 0$ such that either $|u_q (q_k, 1)| \geq \epsilon$ or $u_p (q_k,
1) \leq - \epsilon$ holds for all sufficiently large $k$. Since $u \in C^{1,\alpha}
(\bar S)$, we have
\[ u (q_k \pm t, 1) = u (q_k, 1) \pm u_q (q_k, 1) \, t + O \big( t^{1+\alpha} \big) 
\quad \mbox{as} \ t \to +0 ,
\]
which yields that $\hat u < u (q_k \pm t, 1)$ when $\pm u_q (q_k, 1) > \epsilon$
provided $k$ is sufficiently large and $t>0$ is sufficiently small. However, the
last inequality for $\hat u$ contradicts its definition. Similarly, we have
\[ u (q_k, 1 - t) = u (q_k, 1) - u_p (q_k, 1) \, t + O \big( t^{1+\alpha} \big) 
\quad \mbox{as} \ t \to +0 .
\]
Then the assumption about $u_p (q_k, 1)$ gives that $\hat u < u (q_k, 1 - t)$
provided $t$ is sufficiently small. This is impossible in view of the definition of
$\hat u$. Thus, relations (\ref{eq:12}) are proved.

\subsection{Proof of Theorem 1$'$}

First we prove that $r \geq r_c$ and begin with demonstration that there exists $s
> s_0$ such that $H (1; s) = \check \eta$. For this purpose we consider vorticity
distributions satisfying conditions (i), (ii) and (iii) separately.

If conditions (i) are fulfilled, then $H (1; s)$ decreases monotonically from the
positive infinity to zero as $s$ goes from $s_0$ to $+\infty$. Hence $H (1; s)$
attains the value $\check \eta > 0$ at some $s \in (s_0, +\infty)$.

Since conditions (ii) imply that $s_0 = 0$ and $H_p (0; s) \to +\infty$ as $s \to 0$,
we have that the function $h_p (q,0) - H_p (0; s_*)$ is negative and separated from
zero for some $s_* > 0$. Then $h (q_*, 1) - H (1; s_*) \leq 0$ for some $q_* \in
\RR$ because otherwise the function $h (q, p) - H (p; s_*)$ violates Proposition~1.
It follows from the last inequality that $H (1; s) = \check \eta$ for some $s > s_*$
because $H (1; s)$ monotonically decreases and tends to zero as $s$ grows, whereas
$\check \eta \leq h (q_*, 1)$.

Since conditions (iii) imply that $s_0 > 0$ and $H_p (1; s) \to +\infty$ as $s \to
s_0$, we have that $h_p (q,1) - H_p (1; s_*)$ is negative and separated from zero
for all $q \in \RR$ and some $s_* > s_0$. The latter fact is similar to that
obtained from conditions (ii). This allows us to repeat literally the
considerations used in that case, thus arriving at the conclusion that $H (1; s) =
\check \eta$ for some $s > s_*$.

Let us prove the inequalities $r \geq r_c$ and $h (q, 1) > \check \eta$ for
arbitrary $q$. For this purpose we consider $h (q,p) - H (p; s)$ with $s > s_0$ such
that $H (1; s) = \check \eta$ (the existence of that $s$ was just established).
According to Proposition~2, we have $h_p (q_k, 1) - H_p (1; s) \leq \epsilon_k$ for
some sequences $\{ q_k \}_{k=1}^\infty$ and $\{ \epsilon_k \}_{k=1}^\infty$; the
last of these has positive elements and tends to zero. Combining this and
Bernoulli's equation for $h$, we get
\begin{equation}
r = \frac 1 3 \left[ \frac {1 + h_q^2} {h_p^2} + 2h \right] \geq \frac 1 3 \left[
\frac 1 {(H_p(1,s) + \epsilon_k)^2} + 2h(q_k, 1) \right] . \label{eq:circ}
\end{equation}
Letting $k \to \infty$ in this inequality and taking into account relation
(\ref{eq:11}) and the definition of ${\cal R} (s)$, we obtain that $r \geq {\cal R}
(s)$, and so $r \geq r_c$.

In order to prove the strict inequality $r > r_c$, let us assume that problem
(\ref{eq:9}) with $r = r_c$ has a non-stream solution $h$. To show that this is
impossible we use the equality
\begin{eqnarray}
&& \left[ \Phi (1; s) - 1 \right] \int_{-q_1}^{q_2} w (q, 1; s) \, \D q +
\int_{-q_1}^{q_2} \! \int_0^1 \frac{H_p^2 w_q^2 + (2 h_p + H_p) w_p^2}{2 h_p^2} \,
\D p \D q \nonumber \\ && \ \ \ \ \ \ \ \ \ \ \ \ \ \ \ \ \ \ \ \ \ \ \ \ \ \ \ \ \
\ \ \ \ \ \ \ \ \ \ \ \ \ \ \ \ = - \int_0^1 \left[ \frac{h_q}{h_p} \, \Phi (p; s)
\right]_{q=-q_1}^{q=q_2} \D p \, . \label{eq:wheel}
\end{eqnarray}
Here $H$ is defined by (\ref{eq:10}), $w (q, p; s) = h (q, p) - H (p; s)$ and $\Phi
(p; s) = \int_0^p H_t^3 (t; s) \, \D t$, whereas $q_1, q_2 > 0$. For the derivation
of this equality see the proof of Lemma~3.1 in \cite{Wheel}, where it appears
implicitly. Indeed, the left-hand side of (\ref{eq:wheel}) is similar to the
expression in his formula (3.1). On the other hand, the right-hand side of
(\ref{eq:wheel}) corresponds to the last term in the Wheeler's identity (3.3).

Taking $s = s_c$ in (\ref{eq:wheel}) in which case $\Phi (1; s_c) = 1$ in view of
formula (\ref{eq:10}) for $H$, we reduce (\ref{eq:wheel}) to
\begin{equation}
\int_{-q_1}^{q_2} \! \int_0^1 \frac{H_p^2 w_q^2 + (2 h_p + H_p) w_p^2}{2 h_p^2} \,
\D p \D q = - \int_0^1 \left[ \frac{w_q}{h_p} \, \Phi (p; s_c)
\right]_{q=-q_1}^{q=q_2} \D p \, . \label{eq:red_wheel}
\end{equation} 
Then the positive integral on the left converges as $q_1, q_2 \to +\infty$. This
implies that there exists two sequences $\{ q_1^{(k)} \}_{k=1}^\infty$, $\{
q_2^{(k)} \}_{k=1}^\infty$ that tend to $+\infty$ as $k \to \infty$ and yield the
following relations:
\[ \int_0^1 \Bigg| \frac{w_q \big( q_j^{(k)} , p; s_c \big) }{h_p \big( q_j^{(k)},
p \big)} \Bigg| \Phi (p; s_c) \, \D p \to 0 \quad \mbox{as} \ k \to \infty , \quad j
= 1,2 .
\]
Therefore, we have
\[ \int_{-\infty}^{\infty} \! \int_0^1 \frac{H_p^2 w_q^2 + (2 h_p + H_p) w_p^2}{2 h_p^2}
\, \D p \D q = 0 \, ,
\]
according to which the gradient of $w$ vanishes a.e., and this is incompatible with
the assumption that $h$ is a non-stream solution.

Let us complete the proof of assertion (1). Since there exists $s$ such that $H (1;
s) = \check \eta$ and consequently $r \geq {\cal R} (s)$, we have that $d_- = d
(s_-)$ corresponding to $r$ is less than or equal to $\check \eta$. Indeed, the
latter is equal to $H (1; s) = d (s)$ which decreases monotonically, whereas $s_-
\geq s$ because $r \geq {\cal R} (s)$. Assuming that there exists $q^\circ$ such
that $d_- = h (q^\circ, 1)$, we apply Hopf's lemma which implies that $h_p (q^\circ,
1) - H_p (1; s_-) < 0$ and this is incompatible with (\ref{eq:circ}), where $q =
q^\circ$ instead of $q_k$. The obtained contradiction proves assertion (1).

Let us turn to assertion (2). Taking $r \in (r_c, r_0)$ and assuming that $\check
\eta \geq d_+$ (this is contrary to the right inequality in this assertion), we
apply equality (\ref{eq:wheel}) with $s = s_+$, and so $\Phi (1; s_+) - 1 \geq 0$
now. Therefore, one again obtains a contradiction with the assumption that $h$ is a
non-stream solution by repeating literally the considerations based on
(\ref{eq:wheel}) and used above for proving that $r \neq r_c$. Thus it is shown
that the right inequality holds.

In order to prove the left inequality we again assume the contrary; that is, that
$\hat \eta < d_+$. This implies that there exists $H (p; s)$ such that $H (1; s) =
\hat \eta$. Since for this $H$ the function $w = h - H$ is less than or equal to
zero and its maximum vanishes, Proposition~2 yields that $w (q_k, 1) \to 0$ as $k
\to \infty$ and $w_p (q_k, 1) \geq \epsilon_k > 0$ for some sequence $\{ q_k
\}_{k=1}^\infty$ ($\epsilon_k$ tends to zero as $k \to \infty$). Using this sequence
in Bernoulli's equation, we obtain that $r \leq {\cal R} (s)$, and so either $\hat
\eta \leq d_-$ or $\hat \eta \geq d_+$. The last inequality contradicts to the
assumption made, whereas the former inequality is impossible because $d_- \leq
\check \eta$ (see assertion (1)) and $h$ is a non-stream solution.

Finally, let the vorticity distribution satisfy conditions (iii) and let a
non-stream solution of problem (\ref{eq:9}) corresponding to $r \geq r_0$ exist and
belong to $C^{1, \alpha} (\bar S)$. We note again that conditions (iii) imply
relations $s_0 > 0$ and $H_p (1; s) \to +\infty$ as $s \to s_0$, and so we have that
$h_p (q, 1) - H_p (1; s_*)$ is negative and separated from zero for all $q \in \RR$
and some $s_* > s_0$. Then $h (q, 1) - H (1; s_*) \leq 0$ for all $q \in \RR$
because otherwise the function $h (q, p) - H (p; s_*)$ violates Proposition~2.
Therefore, the last inequality yields that $H (1; s) = \hat \eta$ for some $s >
s_0$. Thus we have that
\[ w (q, p) = h (q, p) - H (p; s) \leq 0 \ \ \mbox{on} \ \bar S \quad \mbox{and} \quad 
\sup_{q \in \RR} h (q, 1) - H (1; s) = 0 .
\]
Applying Proposition~2 to this function, we get a sequence $\{ q_k \}_{k=1}^\infty$
(generally speaking, other than that in the previous paragraph) such that $w (q_k,
1) \to 0$ as $k \to \infty$ and $w_p (q_k, 1) \geq \epsilon_k > 0$ ($\epsilon_k$
again tends to zero as $k \to \infty$). As above, combining this and Bernoulli's
equation, one arrives at the inequality $r \leq {\cal R} (s)$. This inequality is
strict when $r_0 < r$, thus yielding $\hat \eta \leq d_-$, which is impossible. On
the other hand, if $r_0 = r$, then $r = {\cal R} (s)$, and so either $s = s_0$ or
$\hat \eta = d_-$, both of which lead to a contradiction.

The proof is complete.

\subsection{Flows with the vorticity distribution satisfying conditions {\rm (ii)}}

First we examine Stokes waves of small amplitude for $r \geq r_0$. Our aim is to
show that the stream function changes sign within the corresponding flow, and so the
same is true for the horizontal component of velocity. For this purpose we apply the
description of these waves obtained by \cite{KK2} and based on the following two
assumptions:

\vspace{1mm}

\noindent (I) A stream solution $(u (y), d)$ satisfying problem (\ref{eq:ss1}) with
$r > r_c$ is supposed to be such that $u' (d) \neq 0$.

\noindent (II) The dispersion equation $\sigma (\tau) = 0$ corresponding to a stream
solution satisfying assumption (I) has at least one positive root, say, $\tau_0$
such that none of the values $k \tau_0$ $(k = 1,2,\dots)$ is a root of the
dispersion equation. Here
\[ \sigma (\tau) = u' (d) \, \gamma' (d, \tau) - [u' (d)]^{-1} + \omega (1)  \]
and $\gamma (y, \tau)$ solves the boundary value problem
\[ - \gamma'' + [ \tau^2 - \omega' (u) ] \, \gamma = 0 \ \mbox{on} \ (0, d), \quad
\gamma (0, \tau) = 0 , \ \ \gamma (d, \tau) = 1 ,
\]

If $\omega$ belongs to $C^{2, \alpha} (\RR)$ with $\alpha \in (0, 1)$, then
assumptions (I) and (II) guarantee that for all sufficiently small values of the
parameter $t$ there exists a solution to problem (\ref{eq:lapp})--(\ref{eq:bep})
that has the following representation:
\begin{eqnarray}
&& \psi (x,y; t) = u \left( \frac{d \, y}{\eta (x; t)} \right) + t \cos \frac{\tau_0
x}{1 + \lambda (t)} \, W \left( \frac{d \, y}{\eta (x; t)} \right) + o (t) ,
\label{eq:*} \\ && \eta (x; t) = \frac{d}{1 + \lambda (t)} + t \cos \frac{\tau_0 x}{1
+ \lambda (t)} + o (t) . \nonumber
\end{eqnarray}
Here $\lambda (t) \to 0$ as $t \to 0$ and $W$ solves the following problem (see
formulae (38) and (39) in \cite{KK2}):
\begin{equation}
- W'' + [ \tau^2 - \omega' (u) ] \, W = d^{-1} [ y \, u' \, \tau^2 + 2 \, \omega
(u) ] \ \ \mbox{on} \ (0, d) , \quad W (0) = W (d) = 0 ,
\label{eq:nov9}
\end{equation}
and $W' (d) = d^{-1} u' (d) - [u' (d)]^{-1}$.

According to Proposition 3.5\,(ii) in \cite{KK2}, for every $r > r_c$ there exists a
unidirectional flow such that assumption (II) does not hold for the corresponding
stream solution. Besides, if $r > r_0$, then there also exists at least one stream
solution satisfying assumption (II). However, every such solution changes sign on
$(0, d)$ and the same is true for the stream function (\ref{eq:*}) provided $|t|$ is
small. These facts still hold when $r = r_0$ and the stream solution is such that $s
= u' (0) > s_0$.

Let us consider the remaining case, that is, $r = r_0$, $s = s_0$ and $u' (d) \neq
0$. (This occurs when the vorticity distribution satisfies conditions (ii).) To
demonstrate that the stream function (\ref{eq:*}) changes sign we have to check
that $W' (0) \neq 0$. In order to verify this we take $w$ that solves the following
problem:
\[ - w'' + [ \tau^2 - \omega' (u) ] \, w = 0 \ \mbox{on} \ (0, d), \quad
w (0) = 1 , \ \ w (d) = 0 .
\]
Its existence is guaranteed by Lemma 2.2 in \cite{KK2}. Multiplying the first
relation (\ref{eq:nov9}) by $w$, we integrate the result over $(0, d)$ and then
integrate by parts in the same way as in the proof of the cited lemma. This leads to
the equality $W' (0) = d u' (d) w' (d)$ which implies the required inequality.

As in the case when $\omega$ satisfies conditions (iii) we conjecture that only
stream solutions exist when $r \geq r_0$ and conditions (ii) hold for $\omega$ (cf. the
last assertion of Theorem~1$'$). However, if a solution exists for some $r > r_0$,
then we have the following proposition saying that in this case $d_0$ plays the same
role as $d_+$ in assertion (2) of Theorem~1$'$.

\vspace{1mm}

\noindent {\sc Proposition 3.} {\it Let $\omega$ satisfy conditions {\rm (ii)}. If
for some $r > r_0$ problem $(\ref{eq:9})$ has a solution $h \in C^{1, \alpha} (\bar
S)$, then the inequalities $\hat \eta \geq d_0 > \check \eta$ hold for it.}

\vspace{1mm}

{\it Proof.} To prove the left inequality we assume the contrary, that is, $d_0 >
\hat \eta$. Then for some $s > s_0$ there exists a solution of the form
(\ref{eq:10}) such that $H (1, s) = \hat \eta$. Applying Proposition 2 to the
function $w = h - H$ in the same way as for proving the inequality $r \geq {\cal R}
(s)$ in the proof of Theorem~1$'$, we obtain that $d_- \geq \hat \eta$, but
according to assertion (2) of this theorem we have $\check \eta \geq d_-$. Hence $h$
is identically equal to $H$ which is impossible. Thus, the left inequality is
proved.

It was shown in the proof of Theorem 1$'$ that $s > s_0$ can be found from the
equation $H (1; s) = \check \eta$. This implies the right inequality, thus
completing the proof.

\section{Concluding remarks}

We have considered the problem describing steady, rotational, water waves in the
case when no counter-currents are present in a flow of finite depth (such flows are
referred to as unidirectional). It is shown that for the existence of non-stream
solutions the problem's parameter $r$ (Bernoulli's constant/the total head) must be
strictly greater than the critical value $r_c$\,---\,the unique minimum of the
function ${\cal R}$ (see formula (\ref{eq:calR}) for its definition). Stream
solutions describing shear flows with horizontal free surfaces are a kind of trivial
solutions like those describing irrotational uniform flows. Thus, the requirement on
$r$ obtained here for waves with vorticity generalises that proved by \cite{KK1} in
the irrotational case when two uniform {\it conjugate} flows exist for all $r > r_c$
(see formula (13) in \cite{KK11}). Furthermore, another result obtained for
irrotational waves in \cite{KK1} is shown to be also valid for waves with vorticity.
It says that the depths of the pair of conjugate, shear flows serve as bounds for
the supremum and infimum of the wave profile on a unidirectional flow.

According to a unified theory of conjugate flows developed by \cite{Benuni}, their
important feature (apart of being unidirectional) is that they are {\it
transcritical}. This means that one flow is {\it supercritical}, whereas the other
one is {\it subcritical}; that is, the former's (letter's) depth is less (greater)
than that of the critical flow corresponding to $r_c$. (For a given value of the
rate of flow this relates to the opposite relationship between properly defined
values of the flow velocity.) In his paper (see also references cited therein),
\cite{Benuni} characterised the existence of conjugate flows as a common feature for
many hydrodynamic models and emphasised that it `is crucial to the understanding of
observed wave phenomena'.

In this paper the last statement finds the following confirmation. It occurs that
along with $r_c$ another critical value $r_0$ exists in $(r_c, +\infty)$ provided
the vorticity distribution satisfies either of conditions (ii) and (iii) formulated
in \S~1.2. It is known that for $r > r_0$ only the supercritical shear flow is
unidirectional, whereas all other shear flows corresponding to these values of $r$
(at least one such flow exists for every $r > r_0$) have counter-currents (see
\cite{KK3}). Here we proved that only stream solutions describe unidirectional
flows when $r \geq r_0$ and conditions (iii) hold for the vorticity distribution.
This, in particular, implies that the presence of two conjugate flows is essential
for the existence of solitary waves supported by a supercritical shear flow. In the
case when the vorticity distribution satisfies conditions (ii) and $r \geq r_0$ we
demonstrate that all known flows with waves (these are Stokes waves of small
amplitude; see \cite{KK4}) have counter-currents. Therefore, it is reasonable to
conjecture that there are no unidirectional flows with waves for $r \geq r_0$.

There is another reason to refer to $r_0$ as the critical value of the second kind.
Indeed, it is shown in \cite{KK4} that no steady water waves of small amplitude are
supported by a shear flow with a still free surface.

\vspace{2mm}

\noindent {\bf Acknowledgements.} V.~K. and E.~L. were supported by the Swedish
Research Council (VR). N.~K. acknowledges the support from G.\,S.~Magnuson's
Foundation of the Royal Swedish Academy of Sciences and Link\"oping University.

\end{document}